\documentclass[11pt]{article}

\usepackage{amssymb,latexsym}
\usepackage{amsmath,amscd}
\usepackage{theorem}
\usepackage{citesort}
\usepackage{eucal}
\usepackage[latin1]{inputenc}
\usepackage{bbm}

\title{Deformation Quantization: Observable Algebras, States and
  Representation Theory\\[0.5cm]
  {\normalsize Lecture notes for the summer school in Kopaonik, 2002}}
\author{\textbf{Stefan Waldmann}\thanks{E-mail:
    Stefan.Waldmann@physik.uni-freiburg.de}
  \\[0.1cm]
  Fakult{\"a}t f{\"u}r Mathematik und Physik\\
  Albert-Ludwigs-Universit{\"a}t Freiburg\\
  Physikalisches Institut\\
  Hermann Herder Stra{\ss}e 3\\
  D 79104 Freiburg\\
  Germany}
\date{March 2003\\[0.5cm]
  FR-THEP 2003/04}

\renewcommand{\mathbb}[1]{\mathbbm{#1}}

\newcommand{\im}{{\mathrm i}}
\newcommand{\eu}{{\mathrm e}}

\newcommand{\cc}[1]      {\overline{{#1}}}

\newcommand{\id}         {{\mathsf{id}}}
\newcommand{\tr}         {\mathop{{\mathsf{tr}}}}

\newcommand{\End}        {{\mathsf{End}}}
\newcommand{\ring}[1]    {{\mathsf{{#1}}}}

\newcommand{\SP} [1]     {{\left\langle{{#1}}\right\rangle}}

\newcommand{\Unit}       {\mathbb{1}}

\newcommand{\Pol}        {\mathrm{Pol}}

\newcommand{\Rep}[1][{}] {\mathop{\sideset{^*}{_{\mathcal{#1}}}{\operatorname{\textrm{-}\mathrm{Rep}}}}} 
\newcommand{\srep}       {\varrho_{{\scriptscriptstyle\mathrm{S}}}}
\newcommand{\wrep}       {\varrho_{{\scriptscriptstyle\mathrm{Weyl}}}}
\newcommand{\stars}      {\star_{{\scriptscriptstyle\mathrm{S}}}}
\newcommand{\starw}      {\star_{{\scriptscriptstyle\mathrm{Weyl}}}}
\newcommand{\starBCH}    {\star_{{\scriptscriptstyle\mathrm{BCH}}}}

\newtheorem{lemma}{Lemma}

\newtheorem{theorem}[lemma]{Theorem}

\newtheorem{definition}[lemma]{Definition}

\theorembodyfont{\rm} 

\newtheorem{example}[lemma]{Example}

\newtheorem{question}[lemma]{Question}

\begin{document}

\maketitle

\begin{abstract}
    In these lecture notes I give an introduction to deformation
    quantization. The quantization problem is discussed in some detail
    thereby motivating the notion of star products. Starting from a
    deformed observable algebra, i.e. the star product algebra,
    physical applications require to study representations of this
    algebra. I review the recent development of a representation
    theory including techniques like Rieffel induction and Morita
    equivalence. Applications beyond quantization theory are found in
    noncommutative field theories.
\end{abstract}

%
%

\section{Deformation Quantization}
\label{sec:defquant}


\subsection{Motivation: Why quantization?}
\label{sec:motiv}

Let me start with a brief motivation why quantization is still an
important issue in mathematical physics. Of course we know that the
quantum theory of nature provides the more fundamental description
compared to its classical counterpart. Thus quantization is an
artificial problem since nature is already `quantum'. However, it
seems to be very difficult to find \emph{a priori} quantum
descriptions in general. Usually, for us the classical picture is much
easier to obtain. Thus it is reasonable to take this as a starting
point in order to find the more correct quantum description.

Of course we now need some guidelines how to pass from classical
physics to quantum physics. A very vague collection is given by the
\emph{correspondence principle}: 
\begin{itemize}
\item There exists a \emph{classical limit}. The classical description
    is not so bad for a wide range of energies, distances, momenta
    etc.
\item To any classical observable there corresponds a quantum
    observable. There can not be more classical observables as the
    fundamental description is quantum and classical observables only
    arise as classical limits of quantum observables. On the other
    hand, if there are more quantum observables not having a
    reasonable classical limit then `quantization' is hopeless for
    such a system.
\item Classical \emph{Poisson brackets} should correspond to quantum
    commutators of the corresponding observables.
\end{itemize}
Now we have to make these ideas more precise and provide a more
concrete and detailed formulation. In particular one should give
proper definitions of the words `observable', `classical limit', etc.
This turns out to be a much more profound problem than our naive
approach may suggest. It will be useful to separate the problem into
two kind of sub-problems:
\begin{itemize}
\item Generic features and questions which are common to all
    `quantizations'.
\item Specific features and questions depending on the particular
    example.
\end{itemize}
In \emph{deformation quantization} the main emphasize lies on the
generic questions but one can also perform explicit computations for
examples in order to obtain more specific answers. To get an
impression of the complexity of the problem we shall first take a look
at the `generic' classical situation.


\subsection{The classical framework}
\label{sec:classical}

Classical mechanics `lives' on the classical \emph{phase space} $M$
which is the set of all (pure) states of the system or, equivalently,
the set of possible initial conditions for the time development. The
\emph{observables}, i.e. the functions on $M$, have a \emph{Poisson
  bracket}. This indicates that a good mathematical model for $M$ is a
\emph{Poisson manifold}, i.e. a smooth manifold endowed with a smooth
Poisson structure for the smooth functions $C^\infty(M)$ on $M$. Such
a Poisson structure is a bilinear map
\begin{equation}
    \label{eq:PoissonBracket}
    \{\cdot,\cdot\} : C^\infty(M) \times C^\infty(M) \to C^\infty(M)
\end{equation}
which satisfies the conditions:
\begin{itemize}
\item $\cc{\{f, g\}} = \{\cc{f}, \cc{g}\}$ (Reality).
\item $\{f,g\} = - \{g,f\}$ (Antisymmetry).
\item $\{f, gh\} = \{f,g\}h + g \{f,h\}$ (Leibniz rule).
\item $\{f, \{g,h\}\} = \{\{f,g\},h\} + \{g, \{f,h\}\}$ 
      (Jacobi identity).
\end{itemize}
This way the observables $C^\infty(M)$ become a \emph{Poisson
  algebra}. It follows from the linearity and the Leibniz rule that
in local coordinates $x^1, \ldots, x^n$ of $M$ the Poisson bracket is
of the form
\begin{equation}
    \label{eq:LocalPoisson}
    \{f, g\} (x) = 
    \sum_{i,j} \alpha^{ij}(x) 
    \frac{\partial f}{\partial x^i}(x)
    \frac{\partial g}{\partial x^j}(x)
\end{equation}
with local real-valued functions $\alpha^{ij} = - \alpha^{ji}$. Then
the Jacobi identity is equivalent to the following non-linear partial
differential equation
\begin{equation}
    \label{eq:JacobiLocal}
    \sum_{\ell} \left(
        \alpha^{i\ell} \frac{\partial \alpha^{jk}}{\partial x^\ell}
        + \alpha^{j\ell} \frac{\partial \alpha^{ki}}{\partial x^\ell}
        + \alpha^{k\ell} \frac{\partial \alpha^{ij}}{\partial x^\ell}
    \right) = 0.
\end{equation}
The \emph{Poisson tensor} 
$\alpha := \frac{1}{2} \alpha^{ij} \partial_i \wedge \partial_j$ gives
a globally defined $2$-vector field $\alpha \in
\Gamma^\infty(\bigwedge^2 TM)$ on $M$.
\begin{example}[Some Poisson manifolds]
    \label{example:PoissonBracket}
    ~
    \begin{enumerate}
    \item Take $M = \mathbb{R}^{2n}$ with the \emph{canonical Poisson
          bracket}
        \begin{equation}
            \label{eq:CanPoisson}
            \{f,g\} = \sum_{i=1}^n \left( 
                \frac{\partial f}{\partial q^i}
                \frac{\partial g}{\partial p_i}
                -
                \frac{\partial f}{\partial p_i}
                \frac{\partial g}{\partial q^i}
            \right).
        \end{equation}
    \item Let $\mathfrak{g}$ be a Lie algebra with structure constants
        $c^k_{ij}$ and let $\mathfrak{g}^*$ be its dual space with
        linear coordinates $x_1, \ldots, x_n$. Then
        \begin{equation}
            \label{eq:LiePoisson}
            \{f, g\} (x) = \sum_{i,j,k} x_k c^k_{ij} 
            \frac{\partial f}{\partial x_i}(x)
            \frac{\partial g}{\partial x_j}(x)
        \end{equation}
        defines a Poisson bracket on $\mathfrak{g}^*$. 
    \end{enumerate}
\end{example}

A Poisson manifold is called \emph{symplectic} if the matrix
$(\alpha^{ij}(x))$ at any point $x \in M$ is non-degenerate. In this
case
\begin{equation}
    \label{eq:SymplecticForm}
    \omega = \frac{1}{2} \omega_{ij} dx^i \wedge dx^j 
    \quad
    \textrm{with}
    \quad
    (\omega_{ij}) = (\alpha^{ij})^{-1}
\end{equation}
is a closed non-degenerate two-form on $M$, called the
\emph{symplectic form}. Conversely, any such two-form defines a
symplectic Poisson manifold. In the above examples, the first one is
symplectic with the \emph{canonical symplectic form} $\omega = dq^i
\wedge dp_i$ while the second never is symplectic as e.g. the
coefficients of the Poisson tensor vanish at the origin. The theorem
of Darboux states that any symplectic manifold looks locally like the
above example, i.e. one can always find local coordinates $q^i$, $p_i$
such that $\omega = dq^i \wedge dp_i$. Note however, that globally
this needs not to be the case.

Having such a general mathematical framework we have to investigate
which Poisson manifolds actually occur in the daily life of a
physicist. Let me just mention the following examples:
\begin{itemize}
\item $(\mathbb{R}^{2n}, \omega)$ is the well-known phase
    space from classical mechanics.
\item $\mathfrak{g}^*$ with the above Poisson structure always plays a
    role for systems with symmetries described by $\mathfrak{g}$.
\item The phase space of the rigid body is $SO(3) \times
    \mathbb{R}^3$. More generally, cotangent bundles $T^*Q$ describe
    the phase space of a particle moving in the configuration space
    $Q$. There is a canonical symplectic structure on $T^*Q$.
\item For physical systems with constraints, in particular if there are
    gauge degrees of freedom, the \emph{reduced phase space} typically
    has a very complicated Poisson structure.
\end{itemize}
For more examples and further reading I suggest
\cite{abraham.marsden:1985a,cannasdasilva.weinstein:1999a,arnold:1989a}.


\subsection{Canonical quantization and basic examples}
\label{sec:canonical}

Before approaching the general situation let us first recall the
well-known case of canonical quantization on $\mathbb{R}^{2n}$ as well
as other simple examples.

\subsubsection{Canonical quantization on $\mathbb{R}^{2n}$}
\label{sec:canquant}

Let us consider $n=1$ for convenience. Then we assign to the classical
observables $q$ and $p$ the quantum operators $Q = q$ and $P =
-\im\hbar \frac{\partial}{\partial q}$ acting on wave functions
depending on $q$. However, in order to get all interesting observables
quantized we have to specify what should happen to other classical
observables than $q$ and $p$. In particular, we have to specify an
\emph{ordering prescription} for the polynomials in $q$ and $p$ since
$Q$ and $P$ do no longer commute. Here one can chose between several
possibilities, let me just mention two:
\begin{itemize}
\item \emph{Standard ordering}: $\srep(q^np^m) = Q^nP^m$.
\item \emph{Weyl ordering}: $\wrep(q^n p^m)$ is the corresponding
    totally symmetrized polynomial in $Q$ and $P$, e.g.
    \begin{equation}
        \label{eq:WeylExample}
        \wrep(qp^2) = \frac{1}{3}(QP^2 + QPQ + PQ^2).
    \end{equation}
\end{itemize}
One has the explicit formulas for $\srep$ and $\wrep$, which can be
verified easily:
\begin{equation}
    \label{eq:srep}
    \srep(f) = \sum_{r=0}^\infty \frac{1}{r!}
    \left(\frac{\hbar}{\im}\right)^r 
    \left.\frac{\partial^r f}{\partial p^r} \right|_{p=0}
    \frac{\partial^r}{\partial q^r}
\end{equation}
\begin{equation}
    \label{eq:wrep}
    \wrep (f) = \srep(Nf)
    \quad
    \textrm{where}
    \quad
    N = \eu^{\frac{\im\hbar}{2}\Delta}
    \quad
    \textrm{and}
    \quad
    \Delta = \frac{\partial^2}{\partial q \partial p}.
\end{equation}
From these formulas one observes that $\srep$ as well as $\wrep$ can
be extended to all smooth functions of $q$ and $p$ which depend
polynomially on the momentum $p$. These functions shall be denoted by
$\Pol(T^*\mathbb{R})$. Then the maps $\srep$, $\wrep$
provide \emph{linear isomorphisms}
\begin{equation}
    \label{eq:repisodiffops}
    \srep, \wrep: \Pol (T^*\mathbb{R}) \to \mathrm{Diffops} (\mathbb{R}),
\end{equation}
where $\mathrm{Diffops} (\mathbb{R})$ denotes the space of
differential operators with smooth coefficients acting on functions
depending on $q$.

The \emph{main idea} is now to pull-back the noncommutative product of
the differential operators via these quantization maps to obtain a
\emph{new, noncommutative product} for the classical observables
$\Pol(T^*\mathbb{R})$. Thus one defines the \emph{standard ordered}
and the \emph{Weyl ordered star product}, also called the \emph{Moyal
  star product}
\begin{equation}
    \label{eq:starsdef}
    f \stars g = \srep^{-1} (\srep(f) \srep(g)),
\end{equation}
\begin{equation}
    \label{eq:starwdef}
    f \starw g = \wrep^{-1} (\wrep(f) \wrep(g)).
\end{equation}
With some little computation one obtains the following explicit
formulas
\begin{equation}
    \label{eq:starsexpl}
    f \stars g 
    = \sum_{r=0}^\infty \frac{1}{r!}
    \left(\frac{\hbar}{\im}\right)^r
    \frac{\partial^r f}{\partial p^r} 
    \frac{\partial^r g}{\partial q^r}
\end{equation}
\begin{equation}
    \label{eq:starwexpl}
    \begin{split}
        f \starw g 
        &= N^{-1} (Nf \stars Ng) \\
        &=
        \sum_{r=0}^\infty \frac{1}{r!}
        \left(\frac{\hbar}{2\im}\right)^r
        \sum_{s=0}^r \binom{r}{s} (-1)^{r-s}
        \frac{\partial^r f}{\partial q^s \partial p^{r-s}}
        \frac{\partial^r g}{\partial q^{r-s} \partial p^s},
    \end{split}
\end{equation}
where the operator $N$ is as in (\ref{eq:wrep}).

Let me just mention a few properties of the new products $\stars$ and
$\starw$ which are obvious from the above explicit formulas
\eqref{eq:starsexpl} and \eqref{eq:starwexpl}.
\begin{enumerate}
\item $\star$ is associative since $\mathrm{Diffops}(\mathbb{R})$ is an
    associative algebra.
\item $f \star g = fg \; +$ higher orders in $\hbar$.
\item $f \star g - g \star f = \im \hbar \{f,g\} \; +$ 
    higher orders in $\hbar$.
\item $f \star 1 = f = 1 \star f$.
\item $f \star g = \sum_{r=0}^\infty \hbar^r C_r(f,g)$ with
    bidifferential operators $C_r$.
\end{enumerate}
Furthermore one has $\srep(f)^\dag = \srep (N^2 \cc{f}) \ne
\srep(\cc{f})$ whence the standard ordered quantization is physically
\emph{not} appropriate as it maps classical observables,
i.e. real-valued functions, to not necessarily symmetric
operators. The Weyl ordering has this nice property
\begin{equation}
    \label{eq:wrepInv}
    \wrep(f)^\dag = \wrep(\cc{f}),
\end{equation}
whence physically it is the more reasonable choice. On the level of
the star products this means that
\begin{equation}
    \label{eq:starwInv}
    \cc{f \starw g} = \cc{g} \starw \cc{f},
\end{equation}
while for the standard ordered product no such relation is valid.

Let me finally mention that other orderings like anti-standard,
Wick ordering and anti-Wick ordering lead also to star products having
analogous properties.

\subsubsection{The BCH star product on $\mathfrak{g}^*$}
\label{sec:bch}

Another rather simple example of a star product can be obtained for
the Poisson structure on the dual $\mathfrak{g}^*$ of a Lie algebra
$\mathfrak{g}$ as explained in Example~\ref{example:PoissonBracket}.
The following construction is due to Gutt \cite{gutt:1983a} where she
used this BCH star product to obtain a star product on the cotangent
bundle $T^*G$ of the corresponding Lie group $G$.

Denote by $\Pol^\bullet(\mathfrak{g}^*)$ the space of polynomials on
$\mathfrak{g}^*$ which is known to be graded isomorphic to the
symmetric algebra $\bigvee^\bullet(\mathfrak{g})$ over
$\mathfrak{g}$. Then the Poincar\'e-Birkhoff-Witt theorem states that
the later space is isomorphic to the universal enveloping algebra
$\mathcal{U}(\mathfrak{g})$ as a filtered vector space. An explicit
isomorphism is given by the total symmetrization map
\begin{equation}
    \label{eq:symPBW}
    \sigma: v_1 \vee \cdots \vee v_k 
    \mapsto \frac{(\im\hbar)^k}{k!} 
    \sum_{\pi \in S_k} v_{\pi(1)} \circ \cdots \circ v_{\pi(k)}
\end{equation}
where $\vee$ denotes the symmetric tensor product and $\circ$ the
product in $\mathcal{U}(\mathfrak{g})$. Using this linear isomorphism
one defines a new product
\begin{equation}
    \label{eq:BCHstar}
    f \starBCH g = \sigma^{-1} (\sigma(f) \circ \sigma(g))
\end{equation}
for $f, g \in \Pol(\mathfrak{g}^*)$. Again a rather straightforward
computation shows that this product deforms the pointwise commutative
product and gives in the first order of $\hbar$ of the commutator the
Poisson bracket (\ref{eq:LiePoisson}). In order to obtain a more
explicit formula for $\starBCH$ we extend it to exponential functions
$\eu_x(\xi) = \eu^{\xi(x)}$, where $x \in \mathfrak{g}$ and $\xi \in
\mathfrak{g}^*$. One obtains
\begin{equation}
    \label{eq:exeyebch}
    \eu_x \starBCH \eu_y 
    = \eu_{\frac{1}{\im\hbar} \mathrm{BCH}(\im\hbar x, \im\hbar y)},
\end{equation}
where $\mathrm{BCH}(x,y) = x + y + \frac{1}{2}[x,y] + \cdots$ is the
\emph{Baker-Campbell-Hausdorff series}. While the above formula
perfectly makes sense for small $x,y \in \mathfrak{g}$ the convergence
of the BCH series can not be guaranteed for all elements in
$\mathfrak{g}$. Thus in general (\ref{eq:exeyebch}) can only be
understood as a \emph{formal power series in $\hbar$}.

This turns out to be a general problem: The naive extension of
$\stars$, $\starw$ or $\starBCH$ to all smooth functions on the
underlying phase space is \emph{not} possible. In general, the series
in $\hbar$ diverges. However, we have seen that there are usually
`nice' subalgebras where the series converge or even terminate after
finitely many powers in $\hbar$.

On the other hand, on a generic phase space $M$ there is \emph{no}
distinguished Poisson subalgebra of $C^\infty(M)$. Thus we will look
for deformed products only in the sense of formal power series and
study their generic properties. Only \emph{after specifying} a
concrete example one can start to look for a suitable subalgebra where
the formal series actually converge. The choice of a subalgebra
usually requires additional and more specific information than just
the knowledge of the Poisson manifold $M$.


\subsection{General definitions and first results}
\label{sec:definition}

The framework for the general situation will be a Poisson manifold $M$
as classical phase space and the smooth functions $C^\infty(M)$ on $M$
as classical observable algebra. Then the definition of a star product
according to Bayen et al. \cite{bayen.et.al:1978a} generalizes the
properties of $\stars$, $\starw$, $\starBCH$, see also
\cite{dito.sternheimer:2002a,gutt:2000a,weinstein:1994a} for recent
reviews.
\begin{definition}
    \label{definition:starproduct}
    A star product $\star$ for $M$ is an associative
    $\mathbb{C}[[\lambda]]$-bilinear product for
    $C^\infty(M)[[\lambda]]$,
    \begin{equation}
        \label{eq:starproduct}
        f \star g = \sum_{r=0}^\infty \lambda^r C_r(f,g),
    \end{equation}
    such that
    \begin{enumerate}
    \item $f \star g = fg + \cdots$,
    \item $f \star g - g \star f = \im\lambda\{f,g\} + \cdots$,
    \item $f \star 1 = f = 1 \star f$,
    \item $C_r$ is a bidifferential operator.
    \end{enumerate}
    If in addition $\cc{f \star g} = \cc{g} \star \cc{f}$ then the
    star product is called Hermitian.
\end{definition}
The \emph{formal parameter} $\lambda$ corresponds to Planck's constant
$\hbar$ as soon as one has established convergence of the product for
some suitable subalgebra.

If $S = \id + \sum_{r=1}^\infty \lambda^r S_r$ is a formal series of
differential operators $S_r$ on $M$ and if $\star$ is a star product
then
\begin{equation}
    \label{eq:newstar}
    f \star' g = S^{-1} (Sf \star Sg)
\end{equation}
turns out to be again a star product. We have already seen in
(\ref{eq:starwexpl}) that $\stars$ and $\starw$ are related by the
operator $N$ exactly in this way.
\begin{definition}
    \label{definition:equivalence}
    Two star products $\star$ and $\star'$ are called equivalent if
    there is a formal series of differential operators $S = \id +
    \sum_{r=1}^\infty \lambda^r S_r$ such that (\ref{eq:newstar}) holds.
\end{definition}
It is clear that the above relation indeed defines an equivalence
relation. In particular, $\stars$ and $\starw$ are equivalent.
However, the `equivalence' is first of all only a mathematical one
whence we have the following (still not completely answered) question:
\begin{question}
    Which physical properties of `quantization' do only depend on the
    equivalence class $[\star]$ of the star product and which
    properties depend on the specific choice of $\star$?
\end{question}

Before we continue discussing the framework of deformation
quantization and its physical interpretation let me just mention a few
of the strong results obtained in the last two decades:

The \emph{existence} of star products on symplectic phase spaces was
first shown by DeWilde and Lecomte \cite{dewilde.lecomte:1983b} in
1983, Fedosov \cite{fedosov:1986a} gave an independent proof in 1986,
and Omori, Maeda, Yoshioka \cite{omori.maeda.yoshioka:1991a} gave a
third proof in 1991. Gutt \cite{gutt:1983a} considered the example of
linear Poisson structures, i.e. the case of $\mathfrak{g}^*$, in 1983.
The existence of star products in the general Poisson case turned out
to be a much harder problem and it was eventually solved by Kontsevich
\cite{kontsevich:1997:pre} in 1997. Soon after, Cattaneo and Felder
\cite{cattaneo.felder:2000a} gave a TQFT interpretation of
Kontsevich's construction in terms of a Poisson sigma model.

The \emph{classification} of star products up to equivalence was first
achieved by Nest and Tsygan \cite{nest.tsygan:1995a} in 1995 for the
symplectic case and independently by Bertelson, Cahen, and Gutt
\cite{bertelson.cahen.gutt:1997a}, Deligne \cite{deligne:1995a}, as
well as by Weinstein and Xu \cite{weinstein.xu:1998a}. In the general
Poisson case Kontsevich also found the classification
\cite{kontsevich:1997:pre}.

The mathematical framework of deformation quantization is
Gerstenhaber's theory of deformations of associative algebras
\cite{gerstenhaber:1964a,gerstenhaber.schack:1988a}.
As an example from this mathematical background let me mention the
following quite general construction of an associative deformation. In
fact, $\stars$ and $\starw$ are of this form:
\begin{example}[Commuting Derivations]
    \label{example:comder}
    ~
    \\
    Let $\mathcal{A}$ be an associative algebra and let $D_i, E^i:
    \mathcal{A} \to \mathcal{A}$ be pairwise commuting
    derivations. Denote the undeformed product of $\mathcal{A}$ by
    $\mu: \mathcal{A} \otimes \mathcal{A} \to \mathcal{A}$. Then
    \begin{equation}
        \label{eq:CDformula}
        a \star b 
        = \mu \circ \eu^{\lambda \sum_i D_i \otimes E^i} (a \otimes b)
    \end{equation}
    defines an associative product on $\mathcal{A}[[\lambda]]$
    deforming the product of $\mathcal{A}$, see
    \cite[Thm.~8]{gerstenhaber:1968a}.
\end{example}


\subsection{Star products beyond quantization}
\label{sec:beyond}

The mathematical structure of deformed algebras appears not only in
the theory of quantization but also in many other contexts of recent
mathematical physics. In fact, any associative deformation of a
commutative associative algebra is `morally' a star product for some
Poisson bracket. Let me now illustrate this for two examples.


\subsubsection{The quantum plane}
\label{sec:quantumplane}

Consider the two vector fields $x\frac{\partial}{\partial x}$ and
$y\frac{\partial}{\partial y}$ on $\mathbb{R}^2$. A simple computation
shows that they commute whence we can apply the general formula of
Example~\ref{example:comder} to obtain a star product for the functions
$C^\infty(\mathbb{R}^2)$, i.e.
\begin{equation}
    \label{eq:QPstar}
    f \star g = \mu \circ 
    \eu^{\im\lambda x\frac{\partial}{\partial x}
      \otimes y\frac{\partial}{\partial y}} (f \otimes g)
\end{equation}
is an associative deformation. The first order commutator specifies
the following \emph{quadratic} Poisson bracket
\begin{equation}
    \label{eq:quadraticP}
    \{f, g\} 
    = xy \left(
        \frac{\partial f}{\partial x}
        \frac{\partial g}{\partial y}
        -
        \frac{\partial f}{\partial y}
        \frac{\partial g}{\partial x}
    \right).
\end{equation}
A straightforward computation gives then the following commutation
relations
\[
x \star y = \sum_{r=0}\frac{(\im\lambda)^r}{r!}xy = \eu^{\im\lambda}
xy
\quad
\textrm{and}
\quad
y \star x = yx
\]
whence
\begin{equation}
    \label{eq:quantumplane}
    x \star y = q \, y \star x
    \quad
    \textrm{with}
    \quad
    q = \eu^{\im\lambda}.
\end{equation}
Thus one recovers the commutation relations for the quantum plane, see
e.g. \cite{kassel:1995a}.


\subsubsection{Noncommutative field theories}
\label{sec:ncft}

Let me now mention very briefly the noncommutative field
theories. Here one considers the following very simple model. Take a
symplectic constant two-form $B = \frac{1}{2}B_{ij} dx^i \wedge dx^j$
on Minkowski space-time $M$ together with the Weyl product $\star$
with respect to $B$ as discussed before. Then $(M, \star)$ can be
considered as a `quantized space time' and one may ask the question
whether there are some interesting models for a field theory on such a
space time. The idea is to replace an ordinary Lagrangean by a
noncommutative counterpart, e.g.
\begin{equation}
    \label{eq:ClassLag}
    \mathcal{L} = \partial_\mu \phi \partial^\mu \phi - m^2 \phi^2 +
    \alpha \phi^4 + \cdots
\end{equation}
will be replaced by
\begin{equation}
    \label{eq:QLag}
    \tilde{\mathcal{L}} = \partial_\mu \phi \star \partial^\mu \phi 
    - m^2 \phi \star \phi 
    + \alpha \phi \star \phi \star \phi \star \phi + \cdots.
\end{equation}
Such models are studied by various groups, see e.g.
\cite{bahns.doplicher.fredenhagen.piacitelli:2003a:pre,jurco.schupp.wess:2000a,schomerus:1999a,seiberg.witten:1999a}
and \cite{szabo:2001a:pre} for a recent review with additional
references.

Describing such a model within the framework of deformation
quantization has several advantages: First one can treat
\emph{arbitrary} Poisson structures and not only constant $B$.
Furthermore, one can formulate gauge theories in this context and even
extend the deformation program to situations where the classical
fields take their values in arbitrary non-trivial vector bundles. The
later requires the notion of a deformation quantization of vector
bundles \cite{bursztyn.waldmann:2000b,waldmann:2001b}, to which I
shall come back in Section~\ref{sec:vector}.

%
%

\section{States and Representations}
\label{sec:states}


\subsection{The notion of positivity: ordered rings}
\label{sec:rings}

Up to now the star product gives us a model for the quantum observable
algebra build out of the classical observables. However, for a complete
description one also needs a notion for \emph{states}. Here a
conceptual problem arises: Usually, states are described by vectors
(or better: rays) in a complex Hilbert space. But how should the
algebra $C^\infty(M)[[\lambda]]$ act on a Hilbert space. The formal
power series in $\lambda$ certainly do not fit very well to the
analytic structure of a Hilbert space. So is this already the point
where one has to impose convergence conditions for a replacement
$\lambda \leadsto \hbar$? As we shall see, this is not yet the case
and there is a completely intrinsic description of states for formal
star products, see e.g. \cite{bordemann.waldmann:1998a,waldmann:2002a}
and references therein.

The guideline for the following will be the theory of $C^*$-algebras,
see e.g. \cite{bratteli.robinson:1987a,kadison.ringrose:1997a}. Here a
state is identified with its expectation value functional $\omega:
\mathcal{A} \to \mathbb{C}$ which is a linear functional on the
observables such that $\omega(A^*A) \ge 0$, i.e. $\omega$ is a
\emph{positive linear functional}.
\begin{example}
    \label{example:PosFunBH}
    Let $\mathcal{A}$ be the algebra $\mathcal{B}(\mathcal{H})$ of
    bounded operators on a Hilbert space $\mathcal{H}$ and let $0 \ne
    \phi \in \mathcal{H}$. Then
    \begin{equation}
        \label{eq:MyFirstState}
        \omega(A) = \frac{\SP{\phi, A\phi}}{\SP{\phi,\phi}}
    \end{equation}
    defines a state for $\mathcal{A}$. More generally, one can also
    consider the \emph{mixed states} $\omega(A) = \tr(\varrho A)$
    where $\varrho$ is a density matrix.
\end{example}

Thus the idea will be to use positive functionals as states for star
products as well. Thus we are looking for functionals
\begin{equation}
    \label{eq:PosFun}
    C^\infty(M)[[\lambda]] \to \mathbb{C}[[\lambda]],
\end{equation}
which are now required to be $\mathbb{C}[[\lambda]]$-linear and
positive in the sense of formal power series:
\begin{definition}
    \label{definition:PosPowerSer}
    A real formal power series $a = \sum_{r=r_0}^\infty \lambda^r
    a_r \in \mathbb{R}[[\lambda]]$ is positive if $a_{r_0} > 0$.
\end{definition}
This definition is now used to make sense out of the requirement
\begin{equation}
    \label{eq:PosFunStarProd}
    \omega (\cc{f} \star f) \ge 0
\end{equation}
for $f \in C^\infty(M)[[\lambda]]$. In fact, it turns
$\mathbb{R}[[\lambda]]$ into an \emph{ordered ring}.
\begin{definition}
    \label{definition:OrderedRing}
    An associative, commutative, unital ring $\ring{R}$ is called
    ordered with positive elements $\ring{P} \subseteq \ring{R}$ if 
    $\ring{P} \cdot \ring{P} \subseteq \ring{P}$, $\ring{P} + \ring{P}
    \subseteq \ring{P}$ and $\ring{R}$ is the disjoint union $\ring{R}
    = -\ring{P} \cup \{0\} \cup \ring{P}$.
\end{definition}
Examples for ordered rings in this sense are $\mathbb{Z}$,
$\mathbb{Q}$, $\mathbb{R}$ and $\mathbb{R}[[\lambda]]$. Moreover, if
$\ring{R}$ is ordered then $\ring{R}[[\lambda]]$ is ordered as well by
an analogous definition as for $\mathbb{R}$.


\subsection{$^*$-Algebras over ordered rings}
\label{sec:algebras}

Since the crucial concept of states in deformation quantization is now
based on the notion of ordered rings I shall continue with some
arbitrary ordered ring and consider $^*$-algebras over ordered rings
in general. This way, I shall develop a representation theory for such
$^*$-algebras parallel to the well-known theory of $C^*$-algebras,
generalizing the latter. The star products will then provide a particular
example.

Thus let $\ring{R}$ be an ordered ring and denote by $\ring{C} =
\ring{R}(\im)$ its ring extension by a square root $\im$ of $-1$, i.e.
$\im^2 = -1$. This way one obtains a replacement for the real and
complex numbers $\mathbb{R}$ and $\mathbb{C}$. For deformation
quantization $\ring{R} = \mathbb{R}[[\lambda]]$ and $\ring{C} =
\mathbb{C}[[\lambda]]$ will be the relevant choice.
\begin{definition}
    \label{definition:PreHilbert}
    A pre Hilbert space $\mathcal{H}$ over $\ring{C}$ is a
    $\ring{C}$-module with inner product $\SP{\cdot,\cdot}:
    \mathcal{H} \times \mathcal{H} \to \ring{C}$ such that
    \begin{enumerate}
    \item $\SP{\phi, z\psi + w \chi} = z \SP{\phi,\psi} + w
        \SP{\phi,\chi}$ for $\phi,\psi,\chi \in \mathcal{H}$ and $z,w
        \in \ring{C}$.
    \item $\SP{\phi,\psi} = \cc{\SP{\psi,\phi}}$ for $\phi, \psi \in
        \mathcal{H}$.
    \item $\SP{\phi,\phi} > 0$ for $\phi \ne 0$.
    \end{enumerate}
\end{definition}
An operator $A: \mathcal{H} \to \mathcal{H}$ is called
\emph{adjointable} if there exists an operator $A^*$ such that
\begin{equation}
    \label{eq:Adjoint}
    \SP{\phi, A\psi} = \SP{A^*\phi, \psi}
\end{equation}
for all $\phi, \psi \in \mathcal{H}$. This allows to define
\begin{equation}
    \label{eq:BH}
    \mathcal{B}(\mathcal{H}) = \{A \in \End(\mathcal{H}) \; | \; 
    A \; \textrm{is adjointable} \; \}.
\end{equation}
Note that for a complex Hilbert space this definition exactly yields
the bounded operators thanks to the Hellinger-Toeplitz theorem, see
e.g.~\cite[p.~117]{rudin:1991a}.
\begin{lemma}
    \label{lemma:BHAlg}
    Let $A, B \in \mathcal{B}(\mathcal{H})$ and $z,w \in \ring{C}$.
    \begin{enumerate}
    \item $A^*$ is unique and $A^* \in \mathcal{B}(\mathcal{H})$,
        $A^{**} = A$.
    \item $zA +wB \in \mathcal{B}(\mathcal{H})$ and $(zA+wB)^* =
        \cc{z}A^* + \cc{w} B^*$.
    \item $AB \in \mathcal{B}(\mathcal{H})$ and $(AB)^* = B^*A^*$.
    \end{enumerate}
\end{lemma}
The algebra $\mathcal{B}(\mathcal{H})$ gives now the proto-type for a
$^*$-algebra over $\ring{C}$:
\begin{definition}
    \label{definition:starAlg}
    An associative algebra $\mathcal{A}$ over $\ring{C}$ is called
    $^*$-algebra if it is equipped with an involutive anti-linear
    anti-automorphism $^*:\mathcal{A} \to \mathcal{A}$, called the
    $^*$-involution. A $^*$-homomorphism $\Phi: \mathcal{A} \to
    \mathcal{B}$ between $^*$-algebras is an algebra morphism with
    $\Phi(A^*) = \Phi(A)^*$.
\end{definition}
In the following I shall mainly focus on \emph{unital} $^*$-al\-ge\-bras
and $^*$-homo\-morphisms are required to be unit preserving.
\begin{example}[$^*$-Algebras]
    \label{example:StarAlgs}
    ~
    \begin{enumerate}
    \item Any $C^*$-algebra is a $^*$-algebra over $\mathbb{C}$.
    \item $(C^\infty(M)[[\lambda]], \star)$ with a Hermitian star
        product is a $^*$-algebra over $\mathbb{C}[[\lambda]]$.
    \item $\mathcal{B}(\mathcal{H})$ is a $^*$-algebra over $\ring{C}$
        for any pre Hilbert space $\mathcal{H}$ over $\ring{C}$.
    \item If $\mathcal{A}$ is a $^*$-algebra then $M_n(\mathcal{A})$
        becomes a $^*$-algebra as well.
    \end{enumerate}
\end{example}
\begin{definition}
    \label{definition:SRepA}
    A $^*$-representation $\pi$ of $\mathcal{A}$ on $\mathcal{H}$ is a
    $^*$-homomorphism $\pi: \mathcal{A} \to \mathcal{B}(\mathcal{H})$.
\end{definition}
Given two $^*$-representations $(\mathcal{H}, \pi)$ and $(\mathcal{K},
\varrho)$ then an \emph{intertwiner} is an adjointable isometric map
$T: \mathcal{H} \to \mathcal{K}$ such that
\begin{equation}
    \label{eq:Intertwiner}
    T \pi(A) = \varrho(A) T
\end{equation}
for all $A \in \mathcal{A}$. Two $^*$-representations are called
\emph{equivalent} if there exists a unitary intertwiner between them.
\begin{definition}
    \label{definition:RepTheory}
    The representation theory of $\mathcal{A}$ is the category
    $\Rep(\mathcal{A})$ of all $^*$-representations of $\mathcal{A}$
    with intertwiners as morphisms.
\end{definition}


\subsection{Positive functionals}
\label{sec:posfun}

Since the observable algebras in deformation quantization are
particular examples of $^*$-algebras over $\ring{C}$ we shall now come
back to the question of states in the general framework.
\begin{definition}
    \label{definition:PosFun}
    Let $\mathcal{A}$ be a $^*$-algebra over $\ring{C}$. A linear
    functional $\omega: \mathcal{A} \to \ring{C}$ is called positive
    if
    \begin{equation}
        \label{eq:posfundef}
        \omega(A^*A) \ge 0
    \end{equation}
    for all $A \in \mathcal{A}$. It is called a state of $\mathcal{A}$
    if in addition $\omega(\Unit) = 1$. Then the value $\omega(A)$ is
    called the expectation value of $A \in \mathcal{A}$ in the state
    $\omega$.
\end{definition}
Thus we have to investigate the positive functionals for star
products. The following examples show that this is indeed a
non-trivial notion:
\begin{example}[Positive functionals \protect{\cite{bordemann.waldmann:1998a}}]
    \label{example:PosFunDeltaSchr}
    ~
    \begin{enumerate}
    \item The $\delta$-functional on $\mathbb{R}^{2n}$ is \emph{not}
        positive with respect to the Weyl star product $\starw$ since
        e.g.
        \begin{equation}
            \label{eq:deltanot}
            \delta(\cc{H} \starw H) = - \frac{\lambda^2}{4} < 0,
        \end{equation}
        where $H$ is the Hamiltonian of the harmonic oscillator.
        
        Thus points in phase space are (in general) no longer states
        in quantum mechnics, which is physically to be expected from
        the uncertainty relations.

    \item Consider $f \in C^\infty_0(\mathbb{R}^{2n})[[\lambda]]$ then
        the Schrödinger functional
        \begin{equation}
            \label{eq:schroedinger}
            \omega(f) = \int_{\mathbb{R}^n} f(q, p=0) \; d^nq
        \end{equation}
        turns out to be a positive functional with respect to the Weyl
        star product $\starw$. In this case the positivity can be
        shown by successive partial integrations.
    \end{enumerate}
\end{example}

Having the above examples in mind the following question naturally
arises: How many positive functionals does the algebra
$(C^\infty(M)[[\lambda]], \star)$ have? Classically one knows that the
positive functionals of $C^\infty(M)$ are precisely the compactly
supported positive Borel measures. In particular, there are `many'
positive functionals as e.g. all the $\delta$-functionals are
positive.
\begin{definition}
    \label{definition:posdef}
    A Hermitian deformation $\star$ of a $^*$-algebra $\mathcal{A}$
    over $\ring{C}$ is called a positive deformation if for any
    positive linear functional $\omega_0: \mathcal{A} \to \ring{C}$
    there exist `quantum corrections' $\omega_r$ such that
    \begin{equation}
        \label{eq:defposfun}
        \omega = \sum_{r=0}^\infty \lambda^r \omega_r:
        \mathcal{A}[[\lambda]] \to \ring{C}[[\lambda]]
    \end{equation}
    is positive with respect to $\star$.
\end{definition}
For star products one has the following characterization:
\begin{theorem}[Positive deformations \protect{\cite{bursztyn.waldmann:2000a}}]
    \label{theorem:posdef}
    Any Hermitian star product on a symplectic manifold is a positive
    deformation.
\end{theorem}
The example (\ref{eq:deltanot}) shows that the `quantum corrections'
are indeed necessary in some cases. Physically speaking, the above
theorem states that every classical state arises as classical limit of
a quantum state. This is to be expected in order to have a consistent
classical limit.


\subsection{The GNS construction}
\label{sec:gns}

The GNS construction will now provide a way how to pass from the
abstractly given observable algebra back to usual operators on a pre
Hilbert space, i.e. it allows to construct a $^*$-representation of
the observable algebra. We start with a $^*$-algebra $\mathcal{A}$
over $\ring{C}$ as before and a positive functional $\omega:
\mathcal{A} \to \ring{C}$. As in the well-known case of $C^*$-algebras
a positive functional is real in the sense that
\begin{equation}
    \label{eq:omegaReal}
    \omega(A^*B) = \cc{\omega(B^*A)}
\end{equation}
and it satisfies the \emph{Cauchy-Schwarz inequality}
\begin{equation}
    \label{eq:CSU}
    \omega(A^*B)\cc{\omega(A^*B)} \le \omega(A^*A) \omega(B^*B)
\end{equation}
for all $A, B \in \mathcal{A}$. Now consider the following
subset, the so-called Gel'fand ideal of $\omega$,
\begin{equation}
    \label{eq:Jomega}
    \mathcal{J}_\omega 
    = \{A \in \mathcal{A} \; | \; \omega(A^*A) = 0\},
\end{equation}
which indeed turns out to be a left ideal of $\mathcal{A}$ thanks
to~\eqref{eq:CSU}. Being a left ideal makes the quotient
\begin{equation}
    \label{eq:Homega}
    \mathcal{H}_\omega = \mathcal{A} \big/ \mathcal{J}_\omega
\end{equation}
a $\mathcal{A}$-left module. The module action is usually denoted by
$\pi_\omega(A) \psi_B = \psi_{AB}$ where $\psi_B \in
\mathcal{H}_\omega$ is the equivalence class of $B \in
\mathcal{A}$. Furthermore, $\mathcal{H}_\omega$ is a pre Hilbert space
over $\ring{C}$ with the inner product
\begin{equation}
    \label{eq:InnerProd}
    \SP{\psi_A, \psi_B} = \omega(A^*B).
\end{equation}
Indeed, $\SP{\cdot,\cdot}$ is well-defined and positive definite,
again according to the Cauchy-Schwarz inequality. The final step in
the GNS construction consists in observing that the module action
$\pi_\omega$ is actually a $^*$-representation, i.e.
\begin{equation}
    \label{eq:piomega}
    \SP{\pi_\omega(A) \psi_B, \psi_C} 
    = \SP{\psi_B, \pi_\omega(A^*)\psi_C},
\end{equation}
which can be verified easily. Thus one obtains a $^*$-representation
$(\mathcal{H}_\omega, \pi_\omega)$ of $\mathcal{A}$, called the
\emph{GNS representation} induced by $\omega$. In some sense the GNS
construction is a generalization of the usual Fock space construction
starting from a vacuum vector. In fact, if $\omega$ is even a state,
i.e. $\omega(\Unit) = 1$, then it can be written as expecation value
\begin{equation}
    \label{eq:omegaExp}
    \omega(A) = \SP{\psi_\Unit, \pi_\omega(A)\psi_\Unit},
\end{equation}
whence the vector $\psi_\Unit$ plays the role of the vacuum vector.
\begin{example}[GNS representations]
    \label{example:GNS}
    ~
    \begin{enumerate}
    \item Let $\phi \in \mathcal{H}$ with $\SP{\phi,\phi} = 1$ and
        consider the positive functional $\omega(A) = \SP{\phi,A\phi}$
        on the $^*$-algebra $\mathcal{B}(\mathcal{H})$. Then
        $\mathcal{J}_\omega = \{A \; | \; A\phi = 0\}$ whence
        \begin{equation}
            \label{eq:GNSREproduction}
            \mathcal{H}_\omega =
            \mathcal{B}(\mathcal{H})\big/\mathcal{J}_\omega \ni \psi_A
            \; \mapsto \; A\phi \in \mathcal{H}
        \end{equation}
        is a unitary isomorphism and in fact an intertwiner. Thus one
        recovers the usual action of $\mathcal{B}(\mathcal{H})$ on
        $\mathcal{H}$ by the GNS construction.
    \item Consider again the Schrödinger functional $\omega$ as in
        (\ref{eq:schroedinger}) which is positive for the Weyl star
        product. Then the GNS representation induced by this $\omega$
        is canonically unitarily equivalent to the usual (formal)
        Schrödinger representation in Weyl ordering $\wrep$ on the pre
        Hilbert space $C^\infty_0(\mathbb{R}^n)[[\lambda]]$ of formal
        wave functions, see e.g. \cite{bordemann.waldmann:1998a}.
    \end{enumerate}
\end{example}
The last example is important as it shows that deformation
quantization has the capability to obtain intrinsically the usual
formulation of quantum mechanics starting from the star product
algebras.

Having the nice construction of representations out of positive
functionals one has to investigate whether there are any of
them. In particular, one can construct (somehow pathological) examples
of $^*$-algebras without \emph{any} positive linear functional, see
e.g. \cite{bursztyn.waldmann:2001b}. This motivates the following
definition:
\begin{definition}
    \label{definition:SuffPos}
    A $^*$-algebra $\mathcal{A}$ over $\ring{C}$ has sufficiently many
    positive linear functionals if for any $0 \ne H = H^* \in
    \mathcal{A}$ there exists a positive functional $\omega$ with
    $\omega(H) \ne 0$.
\end{definition}
Physically speaking, this means that we can `measure' whether an
observable is zero or not by examining all possible expectation
values.

By taking the direct sum of all GNS representations one arrives at the
following theorem:
\begin{theorem}[Faithful $^*$-representations \protect{\cite{bursztyn.waldmann:2001b}}]
    \label{theorem:GNS}
    Let $\mathcal{A}$ be a unital $^*$-al\-ge\-bra. Then $\mathcal{A}$
    has sufficiently many positive linear functionals if and only if
    it has a faithful $^*$-representation. In this case $\mathcal{A}$
    can be viewed as a $^*$-subalgebra of some
    $\mathcal{B}(\mathcal{H})$.
\end{theorem}
\begin{lemma}
    \label{lemma:PosDef}
    Let $(\mathcal{A}[[\lambda]], \star)$ be a positive Hermitian
    deformation. Then $\mathcal{A}$ has sufficiently many positive
    linear functionals iff $(\mathcal{A}[[\lambda]], \star)$ has
    sufficiently many positive linear functionals.
\end{lemma}

In particular, since $C^\infty(M)$ has sufficiently many positive
linear functionals (take e.g. the $\delta$-functionals) it follows
that Hermitian star products also have sufficiently many positive
linear functionals. Hence the star product algebra
$(C^\infty(M)[[\lambda]], \star)$ has a faithful $^*$-representation.
This can also be obtained more directly and explicitly.

Let me finally mention some more applications of the GNS construction
in deformation quantization:
\begin{enumerate}
\item There are analogues of $\stars$ and $\starw$ for any cotangent
    bundle $T^*Q$ such that one has a Schrödinger representation on
    $C^\infty_0(Q)[[\lambda]]$ being the GNS representation for a
    Schrödinger functional consisting of integration over $Q$ with
    respect to some positive density
    \cite{bordemann.neumaier.waldmann:1999a}.
\item The WKB expansion \cite{bates.weinstein:1995a} can be obtained
    in a particular GNS representation where the positive functional
    is a particular integration over the graph of $dS$ in $T^*Q$ where
    $S: Q \to \mathbb{R}$ is a solution of the Hamilton-Jacobi
    equation $H \circ dS = E$, see
    \cite{bordemann.neumaier.waldmann:1999a,bordemann.neumaier.pflaum.waldmann:1998a:pre}.
\item There is also a characterization of thermodynamical states using
    the KMS condition. It turns out that for a given inverse
    temperature $\beta$ and a Hamiltonian $H$ the KMS states are
    unique and of the form
    \begin{equation}
        \label{eq:KMS}
        \omega(f) = \tr\left(\mathrm{Exp}(-\beta H) \star f\right),
    \end{equation}
    where $\mathrm{Exp}$ is the star exponential and $\tr$ the unique
    trace for a symplectic star product. The GNS representation turns
    out to be faithful with commutant being (anti-) isomorphic to the
    algebra itself. This gives a sort of baby-version of the
    Tomita-Takesaki theorem
    \cite{basart.flato.lichnerowicz.sternheimer:1984a,bordemann.roemer.waldmann:1998a,waldmann:2000a}.
\end{enumerate}

%
%

\section{Representation Theory}
\label{sec:rep}

As we have seen, formal star products have a rich and physically
relevant representation theory. The main advantage of deformation
quantization becomes transparent when investigating representations:
the algebra of observables is the fundamental object describing the
physical system while the representations depend on the current
situation of the physical system (pure, thermodynamical etc.)  and are
thus a derived concept. The algebra determines its states and
representations but not vice versa. This point of view allows to
consider different representations of the same observable algebra. In
particular, one can speak of super selection rules etc. Thus it is of
primary importance to develop tools for investigating
$\Rep(\mathcal{A})$.


\subsection{Rieffel induction}
\label{sec:rieffel}

Rieffel induction was first developed by Rieffel as a tool for
understanding the representation theory of $C^*$-algebras
\cite{rieffel:1974a,rieffel:1974b}. It turns out that these techniques
are much more algebraic and can also be applied in our context. I
shall just sketch the construction and refer to
\cite{bursztyn.waldmann:2001a,bursztyn.waldmann:2002a} for details.

Let two $^*$-algebras $\mathcal{A}$ and $\mathcal{B}$ be given, where
I always assume that they are unital. The positive functionals allow
us to define positive algebra elements: $A \in \mathcal{A}$ is called
\emph{positive} if for all positive $\omega: \mathcal{A} \to \ring{C}$
the expectation value $\omega(A) > 0$ is positive. The positive
elements are denoted by $\mathcal{A}^+$.

Now consider a $\mathcal{B}$-$\mathcal{A}$ bimodule $\mathcal{E}$. A
$\mathcal{A}$-valued completely positive inner product is a map
\begin{equation}
    \label{eq:AvalInProd}
    \SP{\cdot,\cdot} : \mathcal{E} \times \mathcal{E} \to \mathcal{A}
\end{equation}
such that $\SP{\cdot,\cdot}$ is $\ring{C}$-linear in the second
argument, satisfies $\SP{x,y} = \SP{y,x}^*$ and $\SP{x, y \cdot A} =
\SP{x,y} A$ for all $x,y \in \mathcal{E}$ and $A \in \mathcal{A}$, and
the matrix
\begin{equation}
    \label{eq:cp}
    \left(\SP{x_i,x_j}\right) \in M_n(\mathcal{A})^+
\end{equation}
is positive for all $n$ and $x_1, \ldots, x_n \in \mathcal{E}$. The
inner product is \emph{compatible} with the $\mathcal{B}$-left module
structure if in addition $\SP{B \cdot x,y} = \SP{x, B^* \cdot y}$ for
all $x,y \in \mathcal{E}$ and $B \in \mathcal{B}$.
\begin{example}
    \label{example:BiMod}
    The most important example is the free module $\mathcal{E} =
    \mathcal{A}^n$ over $\mathcal{A}$ and $\mathcal{B} =
    M_n(\mathcal{A})$ acts from the left in the usual way. The inner
    product is then defined by
    \begin{equation}
        \label{eq:inprodfree}
        \SP{x,y} = \sum_{i=1}^n x_i^*y_i,
    \end{equation}
    and it is an easy check that this indeed gives a compatible,
    completely positive $\mathcal{A}$-valued inner product.
\end{example}

Now suppose we have a $^*$-representation $(\mathcal{H}, \pi)$ of
$\mathcal{A}$. Then we consider the tensor product
\begin{equation}
    \label{eq:tildeK}
    \widetilde{\mathcal{K}} 
    := \mathcal{E} \otimes_{\mathcal{A}} \mathcal{H},
\end{equation}
where we use the representation $\pi$ to define the
$\mathcal{A}$-tensor product. Clearly, $\widetilde{\mathcal{K}}$ is a
$\mathcal{B}$-left module as $\mathcal{E}$ was a bimodule. Next, one
endows $\widetilde{\mathcal{K}}$ with an inner product by setting
\begin{equation}
    \label{eq:indInnPro}
    \SP{x \otimes \phi, y \otimes \psi} 
    := \SP{\phi, \pi(\SP{x,y})\psi},
\end{equation}
using the $\mathcal{A}$-valued inner product as well as the inner
product of the pre Hilbert space $\mathcal{H}$. It turns out that
(\ref{eq:indInnPro}) extends sesquilinearily to a well-defined inner
product on $\widetilde{\mathcal{K}}$. Moreover, the complete
positivity of the $\mathcal{A}$-valued inner product guarantees that
(\ref{eq:indInnPro}) is at least positive semi-definite. Thus one can
quotient out the (possible) degeneracy space
$\widetilde{\mathcal{K}}^\bot$ and obtains a pre Hilbert space
$\mathcal{K} := \widetilde{\mathcal{K}} \big/
\widetilde{\mathcal{K}}^\bot$. In a final step one observs that the
$\mathcal{B}$-left module structure on $\widetilde{\mathcal{K}}$ is in
fact compatible with the inner product and descends to $\mathcal{K}$
to give a $^*$-representation $\varrho$ of $\mathcal{B}$. This is the
Rieffel induced representation $(\mathcal{K}, \varrho)$ of
$\mathcal{B}$.
\begin{theorem}[Rieffel induction \protect{\cite{bursztyn.waldmann:2001a}}]
    \label{theorem:rieffel}
    Let $\mathcal{E}$ be a $\mathcal{B}$-$\mathcal{A}$ bimodule with
    compatible, completely positive $\mathcal{A}$-valued inner
    product. Then the construction of the induced representation gives
    a functor
    \begin{equation}
        \label{eq:rieffelind}
        \mathsf{R}_{\mathcal{E}}: 
        \Rep(\mathcal{A}) \ni (\mathcal{H}, \pi) 
        \; \mapsto \;
        (\mathcal{K}, \varrho) \in \Rep(\mathcal{B}).
    \end{equation}    
\end{theorem}
The only thing which remains to be shown is the functoriality whence
we have to specify how $\mathsf{R}_{\mathcal{E}}$ should act on
morphisms. Thus let $U: \mathcal{H}_1 \to \mathcal{H}_2$ be an
isometric adjointable intertwiner. Then $V (x \otimes \phi) := x
\otimes U\phi$ turns out to define an intertwiner between the
$\mathcal{B}$-representations on $\mathcal{K}_1$ and $\mathcal{K}_2$.


\subsection{Strong Morita equivalence}
\label{sec:morita}

Morita equivalence deals with the question under which conditions the
categories of representations are actually equivalent in the sense of
categories. In our situation we have to investigate the categories
$\Rep(\mathcal{A})$ and $\Rep(\mathcal{B})$ for two $^*$-algebras
$\mathcal{A}$ and $\mathcal{B}$. We are interested whether the Rieffel
induction using a bimodule $\mathcal{E}$ gives the equivalence. Since
up to now the algebras $\mathcal{A}$ and $\mathcal{B}$ enter quite
asymmetric we first have to endow the bimodule with some more
structures. Thus we also ask for a $\mathcal{B}$-valued inner product
denoted by
\begin{equation}
    \label{eq:Theta}
    \mathcal{E} \times \mathcal{E} \ni (x,y)
    \; \mapsto \; \Theta_{x,y} \in \mathcal{B},
\end{equation}
which should now be $\ring{C}$-linear and $\mathcal{B}$-linear in the
\emph{first} argument as $\mathcal{E}$ is a $\mathcal{B}$-\emph{left}
module. Beside that it also should be completely positive. Moreover,
the compatibility with the $\mathcal{A}$-module structure is
$\Theta_{x, y \cdot A} = \Theta_{x \cdot A^*, y}$. Finally, the two
inner products $\SP{\cdot,\cdot}$ and $\Theta_{\cdot,\cdot}$ are
called compatible if
\begin{equation}
    \label{eq:ThetaSP}
    \Theta_{x,y} \cdot z = x \cdot \SP{y,z}.
\end{equation}
Given such an $\mathcal{E}$ we can pass to the complex-conjugate
bimodule $\cc{\mathcal{E}}$ which is $\mathcal{E}$ as additive group
but the $\ring{C}$-module structure is now given by $\alpha \cc{x} =
\cc{\cc{\alpha} x}$, where $\mathcal{E} \ni x \mapsto \cc{x} \in
\cc{\mathcal{E}}$ is the identity map. Indeed, one obtains a
$\mathcal{A}$-$\mathcal{B}$ bimodule by the definitions $A \cdot
\cc{x} := \cc{x \cdot A^*}$ and $\cc{x} \cdot B := \cc{B^* \cdot x}$.
Thus the role of $\mathcal{A}$ and $\mathcal{B}$ is exchanged whence
we now have a Rieffel induction functor going the opposite way, namely
$\mathsf{R}_{\cc{\mathcal{E}}}: \Rep(\mathcal{B}) \to
\Rep(\mathcal{A})$.

However, in general this does not yet give an equivalence of
categories, since we do not have made any non-triviality requirements
for the inner products. In principle they can all be chosen to be
identically zero. One says that the inner products are \emph{full} if
\begin{gather}
    \label{eq:FullA}
    \ring{C}\textrm{-span} \{\SP{x,y} \; | \; x,y \in \mathcal{E} \} =
    \mathcal{A} \\
    \label{eq:FullB}
    \ring{C}\textrm{-span} \{\Theta_{x,y} \; | \; x,y \in \mathcal{E} \} =
    \mathcal{B}.
\end{gather}
\begin{definition}[Strong Morita equivalence \protect{\cite{bursztyn.waldmann:2001a}}]
    \label{definition:morita}
    Two unital $^*$-al\-ge\-bras $\mathcal{A}$, $\mathcal{B}$ are
    called strongly Morita equivalent if one finds a bimodule
    $\mathcal{E}$ with compatible completely positive inner products
    which are both full.
\end{definition}
Strong Morita equivalence implies that the algebras behave in many
aspects quite similar. In particular the Rieffel induction functors
$\mathsf{R}_{\mathcal{E}}$ and $\mathsf{R}_{\cc{\mathcal{E}}}$ give an
equivalence of the representation theories $\Rep(\mathcal{A})$ and
$\Rep(\mathcal{B})$. Strong Morita equivalence also implies
(ring-theoretical) Morita equivalence.  Hence the algebras
$\mathcal{A}$ and $\mathcal{B}$ have all the Morita invariants in
common, like isomorphic centers, Hochschild cohomology and many more,
see e.g.~\cite{lam:1999a}. However, there are more invariants, see
\cite{bursztyn.waldmann:2001b}.

One important consequence for the following is that the bimodule
$\mathcal{E}$ viewed as $\mathcal{A}$-right module is always
\emph{finitely generated and projective}. The same holds if we view
$\mathcal{E}$ as a $\mathcal{B}$-left module.


\subsection{An application: deformed vector bundles}
\label{sec:vector}

If we want to understand strong Morita equivalence of star products we
have to investigate the finitely generated projective modules for the
star product algebras. Classically, the finitely generated projective
modules of $C^\infty(M)$ are precisely the sections $\Gamma^\infty(E)$
of some vector bundle $E \to M$. This is the famous Serre-Swan
theorem, see e.g. \cite{swan:1962a}. In Connes' noncommutative
geometry this observation leads to the idea that in a noncommutative
context vector bundles have to be replaced by projective modules
\cite{connes:1994a}.

Thus we are looking for the deformed analog of a vector bundle: In
fact, it is more than an analog as one can easily show that a
projective module over an algebra $\mathcal{A}$ can \emph{always} be
deformed into a projective module over a deformation
$(\mathcal{A}[[\lambda]],\star)$, since one can deform projections.
Here one even has an explicit formula, see
\cite[Eq.~(6.1.4)]{fedosov:1996a}. If $P_0$ is the classical
projection then
\begin{equation}
    \label{eq:defProj}
    \mathbf{P} = \frac{1}{2} + \left(P_0 - \frac{1}{2}\right) 
    \star \frac{1}{\sqrt[\star]{1 + 4 (P_0 \star P_0 - P_0)}}
\end{equation}
defines a projection for the deformed product, i.e. $\mathbf{P} \star
\mathbf{P} = \mathbf{P}$, as a simple computation shows. Based on this
observation one can prove the following statement:
\begin{theorem}[Deformed vector bundles \cite{bursztyn.waldmann:2000b}]
    \label{theorem:defvec}
    Let $\star$ be a star product on $M$ and let $E \to M$ be a vector
    bundle. Then there exists a right module structure $\bullet$ on
    $\Gamma^\infty(E)[[\lambda]]$ for $(C^\infty(M)[[\lambda]],
    \star)$ deforming the usual $C^\infty(M)$-module structure.
    Moreover, $\bullet$ is unique up to equivalence and induces a
    deformation $\star'$ for $\Gamma^\infty(\End(E))[[\lambda]]$ which
    is unique up to isomorphism. Finally, also Hermitian fibre metrics
    can be deformed in a unique way up to isometry.
\end{theorem}

This theorem has at least two important consequences. On one hand it
leads to the \emph{classification of star products up to strong Morita
  equivalence}, see
\cite{bursztyn.waldmann:2002a,jurco.schupp.wess:2002a}. In particular,
if the phase space is a cotangent bundle and a background magnetic
field is present, then Morita equivalence of the star product with
magnetic field switched off and the star product taking the magnetic
field into account means that the \emph{Dirac quantization condition}
for the magnetic charges is fulfilled. Moreover, under the Rieffel
induction which implements the equivalence, the analog of the usual
Schrödinger representation (\ref{eq:wrep}) is mapped to a
representation on the sections of the line bundle (generalized wave
functions) over the configuration space corresponding to the magnetic
field which serves as curvature two-form of this line bundle. Hence
Morita equivalence has a very natural physical interpretation in
quantization theory.

On the other hand, deformed vector bundles provide a playground for
noncommutative field theories.  As already discussed, in a classical
field theory the fields can take their values in a non-trivial vector
bundle and a priori it is not clear what the good noncommutative
analog should be.  But the deformed vector bundle gives exactly what
is needed and, even more important, there is no obstruction for it's
construction. In fact this construction can be made quite explicit
using e.g. an adapted version of Fedosov's construction of a star
product \cite{waldmann:2002b} or Kontsevich's formality in the more
general case of Poisson manifolds but only for line bundles
\cite{jurco.schupp.wess:2002a}. Also, a local description by deformed
transition maps can be obtained
\cite{waldmann:2002a,jurco.schupp.wess:2002a}.
  
%
%

\section*{Acknowledgments}

I would like to thank the organizers of the summer school in Kopaonik
as well as the participants for a wonderful atmosphere and many
interesting discussions.
                             
%
%

\begin{footnotesize}
\renewcommand{\arraystretch}{0.5}

\end{footnotesize}
\end{document}